\documentclass[12pt]{article}

\usepackage{latexsym,amsmath,amssymb}
\usepackage{graphicx}

\usepackage[square, sort&compress, numbers, comma]{natbib}

\newcounter{subequation}[equation]
\makeatletter

\def\bcite{\@ifnextchar [{\@tempswatrue\@bcitex}{\@tempswafalse\@bcitex[]}}
\def\@bcitex[#1]#2{\if@filesw\immediate\write\@auxout{\string\citation{#2}}\fi
  \let\@bcitea\@empty
  \@bcite{\@for\@bciteb:=#2\do
    {\@bcitea\def\@bcitea{,\penalty\@m\ }%
     \def\@tempa##1##2\@nil{\edef\@bciteb{\if##1\space##2\else##1##2\fi}}%
     \expandafter\@tempa\@bciteb\@nil
     \@ifundefined{b@\@bciteb}{{\reset@font\bf ?}\@warning
       {Citation `\@bciteb' on page \thepage \space undefined}}%
     \hbox{\csname b@\@bciteb\endcsname}}}{#1}}
\def\@bcite#1#2{{#1\if@tempswa , #2\fi}}

\def\thesubequation{\theequation\@alph\c@subequation}
\def\@subeqnnum{{\rm (\thesubequation)}}
\def\slabel#1{\@bsphack\if@filesw {\let\thepage\relax
   \xdef\@gtempa{\write\@auxout{\string
      \newlabel{#1}{{\thesubequation}{\thepage}}}}}\@gtempa
   \if@nobreak \ifvmode\nobreak\fi\fi\fi\@esphack}
\def\subeqnarray{\stepcounter{equation}
\let\@currentlabel=\theequation\global\c@subequation\@ne
\global\@eqnswtrue
\global\@eqcnt\z@\tabskip\@centering\let\\=\@subeqncr
$$\halign to \displaywidth\bgroup\@eqnsel\hskip\@centering
  $\displaystyle\tabskip\z@{##}$&\global\@eqcnt\@ne
  \hskip 2\arraycolsep \hfil${##}$\hfil
  &\global\@eqcnt\tw@ \hskip 2\arraycolsep
  $\displaystyle\tabskip\z@{##}$\hfil
   \tabskip\@centering&\llap{##}\tabskip\z@\cr}
\def\endsubeqnarray{\@@subeqncr\egroup
                     $$\global\@ignoretrue}
\def\@subeqncr{{\ifnum0=`}\fi\@ifstar{\global\@eqpen\@M
    \@ysubeqncr}{\global\@eqpen\interdisplaylinepenalty \@ysubeqncr}}
\def\@ysubeqncr{\@ifnextchar [{\@xsubeqncr}{\@xsubeqncr[\z@]}}
\def\@xsubeqncr[#1]{\ifnum0=`{\fi}\@@subeqncr
   \noalign{\penalty\@eqpen\vskip\jot\vskip #1\relax}}
\def\@@subeqncr{\let\@tempa\relax
    \ifcase\@eqcnt \def\@tempa{& & &}\or \def\@tempa{& &}
      \else \def\@tempa{&}\fi
     \@tempa \if@eqnsw\@subeqnnum\refstepcounter{subequation}\fi
     \global\@eqnswtrue\global\@eqcnt\z@\cr}
\let\@ssubeqncr=\@subeqncr
\@namedef{subeqnarray*}{\def\@subeqncr{\nonumber\@ssubeqncr}\subeqnarray}
\@namedef{endsubeqnarray*}{\global\advance\c@equation\m@ne%
                           \nonumber\endsubeqnarray}

\renewcommand\maketitle{\par
  \begingroup
    \if@twocolumn
      \ifnum \col@number=\@ne
        \@maketitle
      \else
        \twocolumn[\@maketitle]%
      \fi
    \else
      \newpage
      \global\@topnum\z@   
      \@maketitle
    \fi
    \thispagestyle{plain}\@thanks
  \endgroup
  \setcounter{footnote}{0}%
  \global\let\thanks\relax
  \global\let\maketitle\relax
  \global\let\@maketitle\relax
  \global\let\@thanks\@empty
  \global\let\@author\@empty
  \global\let\@date\@empty
  \global\let\@title\@empty
  \global\let\title\relax
  \global\let\author\relax
  \global\let\date\relax
  \global\let\and\relax
}

\makeatother

\DeclareFontFamily{OT1}{rsfs11}{}
\DeclareFontShape{OT1}{rsfs11}{m}{n}{ <-> rsfs11 }{}
\DeclareMathAlphabet{\mathscript}{OT1}{rsfs11}{m}{n}

\numberwithin{equation}{section}

\newcommand{\gtlt}{\mathrel{\raise2.5pt\hbox{\oalign{$\scriptstyle>$\crcr
$\scriptstyle<$}}}}

\newcommand{\bo}{\raise-0.4mm\hbox{$\Box$}}              

\newcommand{\be}{\begin{equation}}
\newcommand{\ee}{\end{equation}}
\renewcommand{\[}{\begin{equation}}
\renewcommand{\]}{\end{equation}}

\newcommand{\bea}{\begin{eqnarray}}
\newcommand{\eea}{\end{eqnarray}}
\newcommand{\bsea}{\begin{subeqnarray}}
\newcommand{\esea}{\end{subeqnarray}}

\renewcommand{\d}{\mathrm{d}}

\begin{document}

\begin{titlepage}
\begin{flushright}
{\small
DAMTP-2007-14 \\[-0.5ex]
ITFA-2007-12}

\end{flushright}
\vspace{.5cm}
\begin{center}
\baselineskip=16pt {\LARGE Generating Ekpyrotic Curvature
Perturbations
\\ Before the Big Bang
\\ }
\vspace{10mm}
{ Jean-Luc Lehners$^{1}$, Paul McFadden$^{2}$, Neil Turok$^{1}$
and Paul J. Steinhardt$^{3,4}$} \vspace{15mm}

{\small\it $^1$ DAMTP, CMS, Wilberforce Road, CB3 0WA,
Cambridge, UK
\\ \vspace{6pt}
$^2$ ITFA, Valckenierstraat 65, 1018XE Amsterdam, the Netherlands \\ \vspace{6pt}
$^{3}$ Joseph Henry Laboratories, Princeton University, Princeton, NJ 08544, USA \\
\vspace{6 pt}
$^{4}$ Princeton Center for Theoretical Physics, Jadwin Hall, Princeton University, Princeton, NJ 08544, USA}
\vspace{1cm}

\end{center}

\abstract{We analyze a general mechanism for
producing a nearly scale-invariant spectrum of cosmological
curvature perturbations during a contracting phase preceding a big
bang, that can be entirely described using 4d effective field
theory. The mechanism, based on first producing entropic
perturbations and then converting them to curvature perturbations,
can be naturally incorporated in cyclic and ekpyrotic models in
which the big bang is modelled as a brane collision, as well as
other types of cosmological models with a pre-big bang phase. We
show that the correct perturbation amplitude can be obtained and
that the spectral tilt $n_s$ tends to range from slightly blue to
red, with $0.97 < n_s < 1.02$ for the simplest models, a range
compatible with current observations but shifted by a few per cent
towards the blue compared to the prediction of the simplest,
large-field inflationary models.}

\vspace{2mm} \vfill \hrule width 2.3cm \vspace{2mm}{\footnotesize
\noindent \hspace{-9mm}
 E-mail: \texttt{j.lehners@damtp.cam.ac.uk,
mcfadden@science.uva.nl, n.g.turok@damtp.cam.ac.uk,
steinh@princeton.edu} }
\end{titlepage}

\setcounter{page}{2}

\section{Introduction}

The primordial perturbations so dramatically imaged by WMAP and
other cosmological probes offer some of our best clues as to the
fundamental physics underlying the hot big bang. The observed
perturbations seem extremely simple in character: small-amplitude,
nearly Gaussian, growing-mode perturbations that are
approximately scale-invariant and adiabatic in form. Any proposed
cosmological model has to explain these basic features, and any
predicted deviations from Gaussianity, scale-invariance or
adiabaticity are likely to be critical in allowing us to
discriminate between models.

For over two decades, it has been well-understood that simple
scalar field models of inflation can, with some degree of
fine-tuning, produce density perturbations of the required form.
More recently, an alternative set of cosmological models has been
proposed -- the ekpyrotic and cyclic models
\cite{Khoury:2001wf,Steinhardt:2001st} --
which, it has been argued, also solve the standard
horizon and flatness puzzles before the big bang and generate a nearly
scale-invariant spectrum of
realistic cosmological perturbations.

The ekpyrotic mechanism for generating cosmological perturbations
has been the subject of considerable debate due to a key
conceptual hurdle. In most cosmological models of interest,
including the ekpyrotic and cyclic models, all or most of the
contraction phase prior to the big bang can be described
approximately using 4d effective field theory.   In this
description, the growing-mode, adiabatic perturbations in the
contracting, pre-bang phase have a  geometrical character that is
very different from the  growing-mode adiabatic perturbations in
the expanding, post-bang phase. In an expanding phase, the growing
adiabatic mode is a local dilatation (or curvature perturbation)
of the constant density hypersurfaces. But by time-reversal
invariance, this mode is generically the {\it decaying} mode in
the corresponding contracting universe. Instead, the adiabatic
growing mode in a contracting universe is a time-delay
perturbation to the big crunch which, under time-reversal, maps to
a decaying mode perturbation in an expanding universe.
Consequently, any pre-big bang scenario for generating the
primordial perturbations  must explain how growing mode time-delay
perturbations before the bang convert into growing mode curvature
perturbations after the bang.  This issue has been clearly
analyzed by Creminelli {\it et  al.}~\cite{Creminelli:2004jg}, for
example.

One possibility  is
that it is precisely the breakdown of the 4d
effective theory near the crunch/bang transition that
allows the growing time-delay
mode to transform naturally into a growing mode curvature perturbation.
This possibility has been studied extensively, for example,
in the Randall-Sundrum model
 \cite{Tolley:2003nx,McFadden}, where it was shown
that, near the bang (the collision between branes), the warping of
the 5d bulk causes the 4d effective theory to fail at order
$(V/c)^2$, where $V$ is the brane collision speed. This breakdown
allows for mixing between the growing and decaying modes, which is normally prohibited
within a purely 4d context. To date, the analysis has relied on
matching a 4d effective description far from the bang with a 5d
description near the collision, which is not completely
satisfying. Ultimately, the goal is to describe this mechanism
entirely in 5d, but this requires including the interbrane
potential within a fully 5d calculation, which has not yet been
achieved.  But even if the 5d mechanism is proven rigorously to
work, one may inquire whether these inherently 5d effects are the only approach for obtaining nearly scale-invariant fluctuations during a contracting phase.

In this paper, we show the answer is no:  we analyze here an
alternative ``entropic'' mechanism  -- described entirely in terms
of 4d effective field theory --  that converts growing mode
scale-invariant perturbations developed in a contracting phase
into scale-invariant curvature perturbations just {\it before} the
big crunch/big bang transition. This mechanism, which  has been
suggested previously by Notari and Riotto \cite{Notari:2002yc} (see also Tsujikawa {\it et
al.} \cite{Tsujikawa:2002qc}), uses elements that
are familiar to many cosmologists who study entropic perturbations
in inflationary cosmology. Here we analyze the predictions for the
spectral amplitude and tilt and show that the ingredients needed
to generate entropic perturbations and convert them into curvature
perturbations before the big bang arise naturally in cosmological
models like the ekpyrotic and cyclic picture. The mechanism can
also be applied to other models with a transition from a
contracting to an expanding phase, including those which do not
have a higher dimensional realization. We understand that,
following conference presentations of the work described here by
one of us \cite{NTtalk}, Buchbinder {\it et al.~}\cite{khourynew}
and Creminelli {\it et al.~}\cite{cremnew} have been applying
similar ideas to ekpyrotic models with a non-singular bounce.

In this paper, we shall use as an example heterotic M-theory
\cite{HW1,HW2,LOSW1,LOSW2}, which underlies the ekpyrotic and
cyclic models, and show that it contains in its simplest
consistent truncation all the necessary ingredients. First, there
is not just one scalar modulus, but two. It is quite plausible for
{\it both} of these fields to develop scale-invariant
perturbations in the contracting phase. Since both scalars are
perturbed independently, there is naturally produced a spectrum of
scale-invariant {\it entropy} perturbations in the pre-bang phase.
Second, as has recently been understood \cite{Lehners:2006ir},
there are generically one or more ``hard boundaries'' in moduli
space, at which the 4d scalar field trajectory bounces suddenly
{\it before} the collision between branes (the big crunch). The
resulting bounce from these boundaries turns out to be just what
is required to convert scale-invariant entropy perturbations into
scale-invariant curvature perturbations. Neglecting any
higher-dimensional corrections, the matching prescription of
Ref.~\cite{Tolley:2003nx} then implies that these curvature
perturbations generated just before the bounce
 propagate across the crunch/bang transition and into
growing mode curvature perturbations in the ensuing expanding
phase. Under the assumption that all scalar moduli, with the
possible exception of the radion, are frozen well after the big
bang, the isocurvature perturbations naturally die away, leaving
only the curvature perturbation to seed structure in the late
universe.

The outline of this paper is as follows. Sec.~2 reviews the difference
between time-delay and curvature perturbations and explains why the former naturally dominate in contracting universes.  Sec.~3  reviews the ekpyrotic
mechanism for generating scale-invariant perturbations in scalar fields,
pointing out its wide generality and the connection with classical
scale-invariance.  The analysis is first done without including gravity
in order
 to emphasize that,
unlike the case of inflation, the ekpyrotic mechanism is
essentially non-gravitational. Next, we extend this analysis to
the case of two or more fields and show that it generates an
additional ``entropic'' or isocurvature growing mode. We show that
the perturbations in the absence of gravity generically have a red
tilt of a few percent away from scale-invariant. We then compute
the spectrum of entropy perturbations in the presence of gravity,
showing that there is an additional Planck-scale suppressed, blue
correction to the spectral tilt. Sec.~4 reviews how, if one of
these scalar fields undergoes a sudden change in its evolution
before the bang, it converts the entropic perturbation to a
curvature perturbation.  We discuss how these sudden changes occur
naturally in heterotic M-theory due to various types of ``hard
boundaries'' in scalar field space, in particular the one due to
the bounce of the negative tension brane in heterotic M-theory
\cite{Lehners:2006pu,Lehners:2006ir}. We compute the curvature
perturbation amplitude on long wavelengths and show that it can
match the observed value; for example, if the potential minimum
and other characteristic parameters are set near the unification
scale, the density fluctuation amplitude is ${\cal O}(10^{-5})$.
Sec.~5  focuses on the spectral index. We  present  various
equivalent expressions for the spectral index in terms of: (1) the
equation of state and its time-variation; (2) the scalar field
potential and its derivatives; and, (3) the conventional
parameters $\bar{\epsilon}$ and $\bar{\eta}$
\cite{Gratton:2003pe}. Using both model-independent and
model-dependent analyses, we show that the predicted range of
$n_s$ for the simplest models is between 0.97 and 1.02, a few per
cent bluer than the analogous range for inflationary models or for
ekpyrotic models in which the curvature perturbations are produced
through 5d effects~\cite{Tolley:2003nx,McFadden}. Sec.~6 discusses
the implications of these results.

\section{The Problem of Generating Curvature Perturbations in a Contracting Universe}

To understand the basic difference between the time-delay and
curvature modes, it is helpful to use the following mathematical
trick for generating long-wavelength solutions to the linearized
Einstein equations. One starts by choosing a completely
gauge-fixed gauge, such as conformal Newtonian gauge, according to
which scalar perturbations of a flat FRW line element take the
form \be \d s^2 = -(1+2 \Phi)\,\d t^2 + a^2(t) (1-2 \Psi)\, \d{\bf
x}^2. \label{b1} \ee One way to find the spatial perturbation
modes is to begin with a general linear combination of pure
homogeneous gauge transformations (which, by definition, precisely
satisfy Einstein's equations) and then promote the coefficients to
functions of space. This construction ensures that the combination
automatically satisfies the Einstein equations up to spatial
gradients, which is what we seek.

Gauge transformations, with spatial gradients neglected, of a
homogenous, isotropic background do not generate anisotropic
stress at linear order. In conformal Newtonian gauge, however, as
is well known, the absence of anisotropic stress  implies
$\Psi=\Phi$. It is then straightforward to check that the most
general gauge transformation preserving (\ref{b1}) up to spatial
gradient terms is
 \be \delta t = {\alpha_1({\bf x}) \over a} - {\alpha_2({\bf
x}) \over a} \int^t \d t' a(t'), \qquad \delta x^i =   
\alpha_2({\bf x})\,x^i, \label{b2} \ee
where the constant coefficients have been replaced by functions of space.
Here, $\alpha_1({\bf
x})$ is a local time delay to or from the singularity and
$\alpha_2({\bf x})$ describes a local dilatation (or curvature
perturbation). The resulting Newtonian potential is \be \Phi=
\alpha_1({\bf x}){\dot{a} \over a^2} + \alpha_2({\bf x}) \left(
1- {\dot{a}\over a^2} \int^t \d t' a(t')\right), \ee where dots
denote $t$ derivatives.

In an expanding universe, the former is
the decaying mode and the latter is the growing mode. In a
contracting universe, these roles are reversed. Hence,
the two modes have very different geometrical
characters and, at least within the context of 4d effective
theory, it is not obvious how scale-invariant growing
perturbations developed in a collapsing phase might match across a
crunch/bang transition to growing mode perturbations of the type
required in the ensuing expanding phase. This problem
has been the focus of
some concern amongst those using purely 4d effective
theory to analyze pre-big bang and ekpyrotic
perturbations; see, for example, Ref.~\cite{Creminelli:2004jg}.

As noted in the introduction, one solution to the problem that has been
pursued for ekpyrotic and cyclic models has been the inclusion of
intrinsically 5d effects that have no simple analogue in 4d effective
theory~\cite{Tolley:2003nx,McFadden}.  The remainder of this paper, though, considers a second possibility, an entropic mechanism that may be fully
described within 4d effective theory, that can be applied both to ekpyrotic/cyclic models and to more general cosmological
models with a contracting
phase.

\section{Ekpyrotic Perturbations}

The ekpyrotic mechanism for generating perturbations
is essentially
non-gravitational. It relies on the quantum fluctuations that naturally
occur when scalar fields roll down steep scalar field potentials,
even in the absence of gravity.  The relevant potentials are
similar in form to those which occur naturally in many string theory
and supergravity models. Hence it provides an interesting alternative
to the usual inflationary mechanism.  In this section, we first analyze
the generation of ekpyrotic perturbations of a single scalar field
 in the absence of gravity,
based on a combination of classical scale-invariance and simple quantum
physics.  Then, we extend the mechanism to the case of multiple fields and
point out the production of entropy perturbations. Finally, we consider the
gravitational corrections.

\subsection{Ekpyrotic Perturbations without Gravity}

Consider a scalar field $\phi$ in Minkowski spacetime, with action
\be {\cal S} = \int \d^4 x \left(-{1\over 2} (\partial \phi)^2 +
V_0 e^{-c \phi}\right), \label{scl} \ee where we use signature
$-+++$. We consider the case where $V_0$ is positive so the
potential energy is formally unbounded below and the scalar field
runs to $-\infty$ in a finite time. Note that the actual value of
$V_0$ is not a physical parameter, since by shifting the field
$\phi$ one can alter the value of $V_0$ arbitrarily. Furthermore,
in the ekpyrotic or cyclic models, the potential is expected to
turn up towards zero at large negative $\phi$ (because this
corresponds to the limit in heterotic M-theory in which the string
coupling approaches zero and the potential
disappears~\cite{Steinhardt:2001st}), but this detail is
irrelevant to the generation of perturbations on long wavelengths.

We now argue that, as the background scalar field rolls down the
exponential potential towards $-\infty$, then, to leading order in
$\hbar$, its quantum fluctuations acquire a scale-invariant
spectrum as the result of three features. First, the action
(\ref{scl}) is classically scale-invariant. Second, by re-scaling
$\phi \rightarrow \phi/c$ and re-defining $V_0$, the constant $c$
can be brought out in front of the action and absorbed into
Planck's constant $\hbar$, {\it i.e.}~$\hbar \rightarrow
\hbar/c^2$, in the expression $i {\cal S}/\hbar$ governing the
quantum theory. Finally, it shall be important that $\phi$ has
dimensions of mass in four spacetime dimensions.

To see the classical scale-invariance, note that shifting the
field $\phi \rightarrow \phi +\epsilon$, and re-scaling
coordinates $x^\mu \rightarrow x^\mu e^{c \epsilon/2}$, just
re-scales the action by $e^{c \epsilon}$ and hence is a symmetry
of the space of solutions of the classical field equations. Now we
consider a spatially homogeneous background solution corresponding
to zero energy density in the scalar field. This ``zero energy''
condition is a reasonable initial state to assume for analyzing
perturbations in the cyclic and ekpyrotic models because the phase
in which the perturbations are generated is preceded by a very low
energy density phase with an extended period of accelerated
expansion, like that of today's universe, which drives the
universe into a very low energy, homogeneous state. No new energy
scale enters and the solution for the scalar field is then
determined (up to a constant) by the scaling symmetry:
$\phi_b=(2/c) \ln (-A t)$. Next we consider quantum fluctuations
$\delta \phi$ in this background. The classical equations are
time-translation invariant, so a spatially homogeneous time-delay
is an allowed perturbation, $\phi = (2/c) \ln (-A(t+ \delta t))
\Rightarrow \delta \phi \propto t^{-1}$. On long wavelengths, for
modes whose evolution is effectively frozen by causality, {\it
i.e.} $|kt| \ll 1$, we can expect the perturbations to follow this
behavior. Hence, the quantum variance in the scalar field, \be
\langle \delta \phi^2 \rangle \propto \hbar t^{-2} \label{qv} \ee
Restoring $c$ via $\phi \rightarrow c \phi$ and $\hbar \rightarrow
c^2 \hbar$ leaves the result unchanged. However, since $\delta
\phi$ has the same dimensions as $t^{-1}$ in four spacetime
dimensions, it follows that the constant of proportionality in
(\ref{qv}) is dimensionless, and therefore that $\delta \phi$ must
have a scale-invariant spectrum of spatial fluctuations.

It is straightforward to check this in detail. Setting
$\phi=\phi_b(t)+\delta \phi(t,{\bf x})$, to linear order in
$\delta \phi$ the field equation reads \be \ddot{\delta \phi} = -
V_{,\phi\phi} \,\delta \phi +{\bf \nabla}^2 \delta \phi \label{onef}
\ee Using the zero energy condition for the classical background,
we obtain $V_{,\phi \phi} = c^2 V = - c^2 \dot{\phi_b}^2/2 = -
2/t^2$. Next we set $\delta \phi(t,{\bf x}) = \sum_{\bf k} (a_{\bf
k}  \chi_{k}(t) e^{i {\bf k\cdot x}} + h.c.)$ with $a_{\bf k}$ the
annihilation operator and $\chi_k(t)$ the normalized positive
frequency modes. The mode functions $\chi_k$ obey \be
\ddot{\delta \chi}_{k} ={2\over t^2} \,\chi_{k} -k^2\chi_{k}, \label{chieq} \ee
and the incoming Minkowski vacuum state corresponds to $\chi_k =
e^{- i k t} \left(1-i/(k t)\right)/\sqrt{2 k}$. For large $|kt|$,
this solution tends to the usual Minkowski positive-frequency
mode. But as $|kt|$ tends to zero, each mode enters the growing
time-delay solution described above, with $\chi_k \propto t^{-1}$.
Computing the variance of the quantum fluctuation and subtracting
the usual Minkowski spacetime divergence, we obtain \be \langle
\delta \phi^2 \rangle = \hbar \int {k^2 \d k \over 4 \pi^2} {1 \over
k^3 t^2}, \label{a5} \ee where the integral is taken over $k$
modes which have ``frozen in'' to follow the time-delay mode. In
agreement with the general argument above, we have obtained
scale-invariant spectrum of growing scalar field perturbations. To
recap, classical scale-invariance determines the $t$ dependence of
the perturbations, and dimensional analysis then gives a
scale-invariant spectrum in $k$, in three space dimensions.

Let us now generalize the discussion to a theory which is only
{\it approximately} scale-invariant. As we shall see, the main
change is to alter the coefficient $2$ on the right hand side of
(\ref{chieq}) by a small multiplicative correction. We replace the
potential in (\ref{scl}) by \be V=-V_0 e^{-\int \d\phi\, c(\phi)},
\label{vdef} \ee with $c(\phi)$ a slowly-varying function of
$\phi$. From the zero-energy condition, we find the background
solution obeys \be \sqrt{2 V_0} (-t) = \int \d\phi\, e^{\int
\d\phi \,c/2} \label{req}. \ee We now express the integral as an
expansion in derivatives of $c$ with respect to $\phi$, by writing
$e^{ \int \d\phi c/2} = (2/c)(\d/\d\phi) \, e^{ \int \d\phi c/2}$
and integrating by parts, twice: \be \int e^{\int c/2} = {2\over
c}\, e^{\int c/2}\left(1+2{c_{,\phi}\over c^2}\right) -\int 4
\Big({c_{,\phi} \over c^3}\Big)_{\hspace{-0.5mm},\phi} \, e^{\int
c/2}, \label{eeq} \ee and so on. From (\ref{vdef}) we have
$V_{,\phi\phi} = (c^2-c_{,\phi}) V$. Hence, combining (\ref{req})
and (\ref{eeq}), we find \be V_{,\phi\phi} \approx -{2\over t^2}
\left(1+3 {c_{,\phi}\over c^2} \right), \label{vcorr} \ee plus
corrections involving higher numbers of $\phi$ derivatives. (By
dimensions, each derivative is accompanied by an additional power
of $c^{-1}$). Note that our sign convention is that $\phi$ is
rolling towards $-\infty$, so a positive $c_{,\phi}$ means that
$c$ decreases as the contracting phase proceeds.  This is what
occurs naturally in models like the cyclic universe, where the
steeply decreasing potential eventually bottoms out before the
bang.

The correction $3 c_{,\phi}/c^2$ in (\ref{vcorr}) may be treated
in the first approximation as a constant, and one can then compute
the correction to the spectral index as follows. The normalized
positive frequency solution of (\ref{chieq}), with 2 replaced by
$2\left(1+3(c_{,\phi}/ c^2)\right)$, is, up to a constant, the
Hankel function $H_\nu^{(2)}(-kt)$, with $\nu= \frac{3}{2}\,(1 +
\frac{4}{3}\, c_{,\phi}/c^2) $. Using the small-argument expansion of
the Hankel function, the term $k^{-3}$ in (\ref{a5}) becomes
instead $k^{-3(1+ (4/3)c_{,\phi}/c^2)}$. Hence to leading order in
derivatives of $c$, the deviation of the spectral index from
scale-invariance is \be n_s -1= -4 {c_{,\phi}\over c^2}.
\label{dev} \ee The spectrum is red for positive $c_{,\phi}$, the
natural case in the cyclic model, for example.
This index characterizes is the non-gravitational contribution
to the fluctuation spectrum.

\subsection{Ekpyrotic Perturbations with Two Fields}

The preceding discussion without gravity included is easily
generalized to two or more fields. For example, consider two
decoupled fields with a combined scalar potential \be V_{tot}=
-V_1 e^{-\int c_1 \d \phi_1}-V_2 e^{-\int c_2 \d\phi_2},
\label{c1} \ee where $c_1=c_1(\phi_1)$, $c_2 = c_2(\phi_2)$, and
$V_1$ and $V_2$ are positive constants. We consider potentials in
which the $c_i$ are slowly varying and hence the potentials are
locally exponential in form. Furthermore, for simplicity, we focus
on scaling background solutions in which both fields {\it
simultaneously} diverge to $-\infty$.

We are specifically interested in the entropy perturbation, namely
the relative fluctuation in the two fields, defined as follows \be
\delta s \equiv (\dot{\phi}_1\,\delta \phi_2 -
\dot{\phi}_2\,\delta \phi_1)/
\sqrt{\dot{\phi}_1^2+\dot{\phi}_2^2}. \label{c3} \ee Since this
quantity is gauge-invariant under linearized coordinate
transformations, it can be expected to survive with only
Planck-scale suppressed corrections even when gravity is turned
on.

In the absence of gravity, the equation of motion of $\delta s$ is
(see e.g. \cite{Gordon:2000hv}) \be \ddot{\delta s} + \left(k^2 +
V_{ss} +3 \dot{\theta}^2 \right) \delta s = 0, \label{dseq} \ee
where \bea V_{ss} &=& \frac{\dot{\phi_2}^2 V_{,\phi_1 \phi_1}-2
\dot{\phi_1}\dot{\phi_2} V_{,\phi_1 \phi_2} + \dot{\phi_1}^2
V_{,\phi_2 \phi_2}}{\dot{\phi_1}^2 +\dot{\phi_2}^2}, \\[1ex]
\dot{\theta} &=& \frac{\dot{\phi_2} V_{,\phi_1}- \dot{\phi_1}
V_{,\phi_2}} {\dot{\phi_1}^2 +\dot{\phi_2}^2}.
\label{def-thetadot}\eea Here, $\dot\theta$ measures the bending
of the trajectory in scalar field space, $(\phi_1,\phi_2)$. We
want to consider background solutions where both fields run to
large negative values, with $\dot{\phi_2}$ and $\dot{\phi_1}$
remaining comparable in magnitude. For this to be true, the
trajectory should not bend too strongly.

For simplicity we shall only study the easiest case,
$\dot{\theta}=0$, for which the background scalar field trajectory
is a straight line. In this case, we have \be \dot{\phi}_2 =
\gamma \dot{\phi}_1, \label{phirel} \ee with $\gamma$ an arbitrary
constant. From the definition (\ref{def-thetadot}), the two scalar
field potentials are related, and, by  integrating,
we find that up to
an irrelevant constant, we must have \be V_{tot} = V(\phi_1)
+\gamma^2 V(\phi_2/\gamma), \label{gammarel} \ee for some function
$V(\phi)$. The equation of motion (\ref{dseq}) now becomes \be
\ddot{\delta s} + (k^2 + V_{, \phi \phi}) \ \delta s = 0, \label{dseq2} \ee
where we have set $\phi=\phi_1$. This is exactly the same equation
as that governing the fluctuations of a single field (\ref{onef}),
and it is straightforward to check that the background solution is
also just that for the field $\phi$ in the potential $V(\phi)$.
Hence, the analysis of the previous section may be applied without
change, with $c(\phi) \equiv - ({\rm ln} V)_{,\phi}$. Furthermore,
since $\delta s$ is a canonically
normalized field according to its definition in (\ref{c3}), the power spectrum generated from quantum
fluctuations is given, in the scale-invariant case, by the same
expression as that for a scalar field, namely (\ref{a5}). Let us
emphasize that (\ref{phirel}) is {\it not} an attractor solution of
the background equations. In fact, it is precisely the instability of
(\ref{phirel}) which generates approximately scale-invariant
perturbations, according to (\ref{dseq2}). The question of how the
system enters the background solution (\ref{phirel}) is of course
important, but shall not be addressed in this paper.

In the next subsection, we will consider the gravitational
corrections to the entropy perturbation spectrum, which are suppressed by
inverse powers of the Planck mass $M_{Pl}$. Since the only relevant
physical parameter in the actions we have so far discussed is $c$,
which has inverse mass dimensions, we can expect the gravitational
effects to be small when
$c^{-1}/ M_{Pl} \ll 1$, and then the estimates obtained above
should be accurate.

\subsection{Ekpyrotic Perturbations including Gravity}
\label{section-Gravity}

Next we turn our attention to the more realistic setting where
gravity is included. We consider the action for $N$ decoupled
fields interacting only through gravity: \be \int \d^4 x \sqrt{-g}
\,\Big(\, \frac{1}{2}R
 -\frac{1}{2} \sum_{i=1}^{N} (\partial \phi_{i})^2 -\sum_{i=1}^{N}
 V_i(\phi_i)\,\Big), \label{lag}  \ee where we have chosen units in which $8\pi G\equiv M_{Pl}^{-2} = 1$. In a flat
 Friedmann-Robertson-Walker background with line element $\d s^2= -\d t^2 +a^2(t) \d {\bf x}^2$, the scalar field
 and Friedmann equations are given by \be \ddot{\phi}_i + 3H\dot{\phi}_i + V_{i,\phi_i} = 0 \ee
 and \be H^2 =\frac{1}{3} \left[\frac{1}{2}  \sum_i
 \dot\phi_i^{~2}+ \sum_i V_i(\phi_i)
 \right], \ee
where $H=\dot{a}/a$ and $V_{i,\phi_i} = (\partial
V_i/\partial \phi_i)$ with no summation implied. Another useful
relation is \be \dot{H} = - \frac{1}{2}  \sum_i \dot\phi_i^{~2}.
\label{eq-Hdot}\ee

If all the fields have negative exponential potentials
$V_i(\phi_i)=-V_i\, e^{-c_i \phi_i}$ then as is well-known, the
Einstein-scalar equations admit the scaling solution \be a =
(-t)^p, \qquad \phi_i = {2\over c_i} \ln (-A_i t), \qquad V_i =
\frac{2 A_i^2}{c_i^2} , \qquad p= \sum_i {2 \over c_i^2}.
\label{c2} \ee Thus, if $c_i \gg 1$ for all $i$, we have a very
slowly contracting universe with $p \ll 1$.

As before, we focus on the entropy perturbation since this is a
local, gauge-invariant quantity, and on the case of only two
scalar fields. The entropy perturbation equation (\ref{dseq}) in
flat spacetime is replaced (see e.g. \cite{Gordon:2000hv}) by \be
\ddot{\delta s} + 3H\dot{\delta s} + \left(\frac{k^2}{a^2}
  + V_{ss} + 3\dot{\theta}^2 \right) \delta s =
\frac{4 k^2 \dot\theta}{a^2 \sqrt{\dot{\phi}_1^2 +
\dot{\phi}_2^2}}\, \Phi. \label{eq-entropy}\ee Again, for simplicity
we will focus attention on straight line trajectories in scalar
field space. Since $\dot{\theta}=0$, the entropy perturbation is
not sourced by the Newtonian potential $\Phi$ and we can solve the
equations rather simply. We shall assume, as before, that the
background solution obeys scaling symmetry so that
$\dot{\phi}_2=\gamma \dot{\phi}_1$.

It is convenient at this point to continue the analysis in terms
of conformal time $\tau$. Denoting $\tau$ derivatives with primes,
and introducing the re-scaled entropy field \be\delta S = a(\tau)
\,\delta s, \ee Eq. (\ref{eq-entropy}) becomes \be {\delta S}'' +
\left(k^2 -\frac{a''}{a}
  + a^2 V_{,\phi \phi}
 \right) \delta S = 0. \label{eq-entropy-S}
  \ee
The crucial term governing the spectrum of the
perturbations is then
\be \tau^2\,\Big(\,{a''\over a} - V_{,\phi \phi} \,a^2\Big).
\label{potpert} \ee
When this quantity is approximately 2, we will again
get nearly scale-invariant perturbations.

It is customary to define the quantity \be \epsilon \equiv {3\over
2} (1+w) \equiv {\dot{\phi_1}^2+\dot\phi_2^2 \over 2 H^2} =
{(1+\gamma^2) \dot\phi^2 \over 2 H^2}. \label{eps} \ee In the
background scaling solution,
 \be \epsilon = \frac{c^2}{2(1+\gamma^2)}. \label{epsc}
\ee We proceed by evaluating the quantity in (\ref{potpert}) in an
expansion in inverse powers of $\epsilon$ and its derivatives with
respect to $N$,  where $N = \ln (a/a_{end})$, where $a_{end}$ is
the value of $a$ at the end of the ekpyrotic phase. Note that $N$
decreases as the fields roll downhill and the contracting
ekpyrotic phase proceeds.

We obtain the first term in (\ref{potpert}) by differentiating
(\ref{eq-Hdot}), obtaining \be {a''\over a} = 2 H^2 a^2
\Big(1-{1\over 2} \,\epsilon\Big). \label{app} \ee The second term
in (\ref{potpert}) is found by differentiating (\ref{eps}) twice
with respect to time and using the background equations and the
definition of $N$. We obtain \be a^2 V_{,\phi \phi}= - a^2 H^2
\,\Big( 2 \epsilon^2  - 6 \epsilon - {5\over 2}
\,\epsilon_{,N}\Big) +O(\epsilon^0). \label{vpp} \ee Finally, need
to express ${\cal H} \equiv (a'/a) = a H$ in terms of the
conformal time $\tau$. From (\ref{app}) we obtain \be {\cal H}'=
{\cal H}^2(1-\epsilon), \label{hcp} \ee which integrates to \be
{\cal H}^{-1}  = \int_0^\tau \d \tau (\epsilon -1). \label{inthcp}
\ee Now, inserting $1= \d(\tau)/\d\tau$ under the integral and
using integration by parts we can re-write this as \be {\cal
H}^{-1} = \epsilon \tau \left(1 -{1\over \epsilon} - (\epsilon
\tau)^{-1} \int_0^\tau \epsilon' \tau \d \tau\right).
\label{integ} \ee Using the same procedure once more, the integral
in this expression can be written as \be (\epsilon \tau)^{-1}
\int_0^\tau \epsilon' \tau \d \tau = \frac{\epsilon'
\tau}{\epsilon} - (\epsilon \tau)^{-1} \int_0^\tau
\frac{\d}{\d\tau}(\epsilon' \tau)\,\tau \d \tau. \ee Now using the
fact that $\epsilon' = {\cal H} \epsilon_{,N}$, and that to
leading order in $1/\epsilon$, ${\cal H}$ can be replaced by its
value in the scaling solution (with constant $\epsilon$), ${\cal
H}\tau = \epsilon^{-1}$, we can re-write the second term on the
right-hand side as \be - (\epsilon \tau)^{-1}\int_0^\tau
\frac{\d}{\d\tau}(\epsilon' \tau)\,\tau \d \tau = - (\epsilon
\tau)^{-1}\int_0^\tau \d \tau \frac{1}{\epsilon}\,
\Big(\frac{\epsilon_{,N}}{\epsilon}\Big)_{\hspace{-0.5mm},N}, \ee
which shows that this term is of order ${1}/{\epsilon^2}$ and can
thus be neglected. Altogether we obtain \be {\cal H}^{-1}
=\int_0^\tau \d \tau (\epsilon-1) \approx \epsilon \tau \left(1-
{1\over \epsilon}- {\epsilon_{,N}\over \epsilon^2}\right).
\label{appepsint} \ee Using (\ref{app}) and (\ref{vpp}) with
(\ref{appepsint}) we can calculate the crucial term entering the
entropy perturbation equation, \be
\tau^2\,\Big(\hspace{0.5mm}{a''\over a} - V_{,\phi \phi}
\,a^2\Big) = 2 \left(1 - \frac{3}{2 \epsilon} + \frac{3}{4}
\frac{\epsilon_{,N}}{\epsilon^2}\right) \ee As explained in the
discussion preceding equation (\ref{dev}), the deviation from
scale-invariance in the spectral index of the entropy perturbation
is then given by \be \label{tilt1} n_s -1 = \frac{2}{\epsilon } -
\frac{\epsilon_{,N}}{\epsilon^2}. \ee The first term on the
right-hand side is the gravitational contribution, which, being
positive, tends to make the spectrum blue. The second term is the
non-gravitational contribution, which tends to make the spectrum
red. We will return to this expression in
Sec.~\ref{section-index}, after explaining how these entropic
perturbations are naturally converted to curvature perturbations.

\section{Converting Entropy to Curvature Perturbations}

We have shown how an approximately scale-invariant spectrum of
entropy perturbations may be generated by scalar fields in a
contracting universe. In this section, we will discuss how these
perturbations may be converted to curvature perturbations if the
scalar field undergoes a sudden acceleration, and we will estimate
the curvature perturbation amplitude.

As an example, we will consider the common case where the scalar
field trajectory encounters a boundary in moduli space and bounces
off it. Such a bounce was recently found in heterotic M-theory
\cite{Lehners:2006pu,Lehners:2006ir}, when the negative tension
brane bounces off the zero of the bulk warp factor just before the
positive and negative tension branes collide. We refer to those
papers for further details: for the purpose of this paper, all we
need to know is that in the 4d effective description there are two
scalar field moduli, $\phi_1$ and $\phi_2$, living on the
half-plane $-\infty < \phi_1 < \infty$, $-\infty < \phi_2 < 0$.
Furthermore, the cosmological solution of interest is one in which
$\phi_2$ encounters the boundary $\phi_2=0$, and reflects off it
elastically.

Let us assume that an approximately scale-invariant entropy
perturbation $\delta s$ has been generated, as described in
previous sections, in a contracting phase of the universe in which
both $\phi_1$ and $\phi_2$ run down steep negative potentials. The
question we want to address is whether this entropy perturbation
may be converted into a curvature perturbation on large scales,
within the realm of validity of 4d effective theory. It is not
hard to see how a scalar field ``bounce'' off a boundary in moduli
space readily achieves this feat. As is well known (see e.g.
Ref.~\cite{Gordon:2000hv}), defining ${\cal R}$ to be the
curvature perturbation on comoving spatial slices, for $N$ scalar
fields with general K\"{a}hler metric $g_{ij}(\phi)$ on scalar
field space, the linearized Einstein-scalar field equations lead
to
 \be \dot{\cal R} = - {H \over \dot{H}}
\left( g_{ij} {D^2 \phi^i \over D t^2} s^j -{k^2\over a^2}
\Psi\right), \label{e1} \ee where the $N-1$ entropy perturbations
\be s^i= \delta \phi^i - \dot{\phi}^i \,\frac{g_{jk}(\phi)\,
\dot{\phi}^j \delta \phi^k }{ g_{lm}(\phi)\, \dot{\phi}^l
\dot{\phi^m}} \label{e2} \ee are just the components of $\delta
\phi^i$ orthogonal to the background trajectory, and the operator
$D^2/Dt^2$ is just the geodesic operator on scalar field space. In
our case, things simplify because the scalar field space is flat,
so the metric is $g_{ij}=\delta_{ij}$, and $D/Dt$ reduces to an
ordinary time derivative. Considering only two scalar fields, we
have \be s^1 = - \dot\phi_2 \,\delta
s/\sqrt{\dot\phi_1^2+\dot\phi_2^2}, \qquad s^2 = + \dot\phi_1
\,\delta s/\sqrt{\dot\phi_1^2+\dot\phi_2^2}. \ee For a straight
line trajectory in field space, the right-hand side of (\ref{e2})
vanishes even if the entropy perturbation is nonzero. However, if
there is a departure from geodesic motion, the entropy
perturbation directly sources the curvature perturbation. This can
happen when scalar field potentials are present, and we would
expect it generically to occur when the ekpyrotic potentials turn
off, around $t_{end}<t_b$, but this contribution due to the
resulting bending of the scalar field trajectory will in general
be very model-dependent. In contrast, a simple reflection of one
of the two fields (say $\phi_2$) off a boundary (in this case at
$\phi_2=0$), at some time $t_b$, results in a model-independent
contribution which is easily computed.

For simplicity, we assume that the scalar field bounce occurs
after the ekpyrotic potentials are turned off, so that the
universe is kinetic-dominated from the 4d point-of-view. The
scalar field trajectory is $\dot{\phi_2}=-\tilde{\gamma}
\dot{\phi_1}$, for $t<t_b$, and $\dot{\phi_2}=\tilde{\gamma}
\dot{\phi_1}$, for $t>t_b$, with $\dot{\phi_1}$ constant and
negative in the vicinity of the bounce. The bounce leads to a
delta function on the right-hand side of (\ref{e1}), \be
\frac{D^2\phi_2}{Dt^2} = \delta(t-t_b) \,2 \dot\phi_2(t_b^+), \label{e3} \ee
where $t_b$ is the time of the bounce of the negative-tension
brane. As can be readily seen from (\ref{e1}), if the entropy
perturbations already have acquired a scale-invariant spectrum by
the time $t_b$, then the bounce leads to their instantaneous
conversion into curvature perturbations with precisely the same
long wavelength spectrum.

We can estimate the amplitude of the resulting curvature
perturbation by integrating equation (\ref{e2}) using (\ref{e3}).
Since we have assumed the universe is kinetic-dominated at this
time, $H=1/(3t)$. As pointed out earlier, since the entropy
perturbation (\ref{c3}) is canonically normalized, its spectrum is
given by (\ref{a5}) up to non-scale-invariant corrections. This
expression only holds as long as the ekpyrotic behavior is still
underway: the ekpyrotic phase ends at a time $t_{end}$
approximately given by $|V_{min}|=2/(c^2 t_{end}^2)$. After
$t_{end}$, the entropy perturbation obeys $\ddot{\delta s} +
t^{-1} \dot{\delta s} = 0$, which has the solution $\delta s = A +
B \ln(-t).$ Matching this solution to the growing mode solution
$t^{-1}$ in the ekpyrotic phase, one finds that by $t_b$ the
entropy grows by an additional factor of $1+ \ln(t_{end}/t_b)$.
Employing the Friedmann equation to relate $\dot{\phi_2}=
\tilde{\gamma} \dot{\phi_1}$ to $H$, putting everything together
and restoring the Planck mass, we find for the variance of the
spatial curvature perturbation in the scale-invariant case, \be
\langle {\cal R}^2 \rangle = \hbar \,{c^2 |V_{min}|\over 3 \pi^2
M_{Pl}^2} \,{ \tilde{\gamma}^2 \over (1+\tilde{\gamma}^2)^2}
\,\left(1+\ln(t_{end}/t_b)\right)^2 \int {\d k \over k} \equiv
\int {\d k\over k}\, \Delta^2_{\cal R}(k) \label{r5} \ee for the
perfectly scale-invariant case. Notice that the result depends
only logarithmically on $t_b$: the main dependence is on the
minimum value of the effective potential and the parameter $c$.
Observations on the current Hubble horizon indicate $
\Delta^2_{\cal R}(k) \approx 2.2\times 10^{-9} $. Ignoring the
logarithm in (\ref{r5}), this requires $c |V_{min}|^{1\over 2}
\approx 10^{-3} M_{Pl}$, or approximately the GUT scale. This is
of course entirely consistent with the heterotic M-theory setting
\cite{BD}.  Having shown it is straightforward to obtain the right
amplitude, we next consider the spectral index.

\section{Comparing Predictions for the Spectral Index}
\label{section-index}

If the entropic perturbations are suddenly
converted to curvature perturbations, as in the example considered in the
previous section, the curvature perturbations inherit the
spectral tilt given in (\ref{tilt1}). In this section, we
analyze this relation using several techniques and compare the prediction
to the those for curvature perturbations in inflation and
for a cyclic model in which
 the time-delay fluctuations
are converted to curvature perturbations before the bang.

As a first approach, let us consider the model-independent
estimating procedure used in Ref.~\cite{Khoury:2003vb}.  We begin
by re-expressing Eq.~(\ref{tilt1}) in terms of ${\cal N}$, the
number of e-folds before the end of the ekpyrotic phase (where $\d
{\cal N} = (\epsilon-1)  N$ and
 $\epsilon \gg 1$):
\be n_s -1 = \frac{2}{\epsilon} -  \frac{\d \ln \epsilon}{\d {\cal
N}}. \ee This expression is identical to the case of the Newtonian
potential perturbations derived in \cite{Khoury:2003vb}, except
that the first term has the opposite sign. In this expression,
$\epsilon({\cal N})$ measures the equation of state during the
ekpyrotic phase, which must decrease from a value much greater
than unity to a value of order unity in the last ${\cal N}$
e-folds. If we estimate  $\epsilon \approx {\cal N}^{\alpha}$,
then  the spectral tilt is \be n_s-1 \approx \frac{2}{{\cal
N^{\alpha}}} - \frac{\alpha}{{\cal N}}. \ee Here we see that the
sign of the tilt is sensitive to $\alpha$. For nearly exponential
potentials ($\alpha \approx 1$),  the spectral tilt is $n_s
\approx 1+ 1/{{\cal} N} \approx 1.02$, slightly blue, because the
first term dominates.  However, there are well-motivated examples
(see below) in which the equation of state does not decrease
linearly with ${\cal N}$. We have introduced $\alpha$ to
parameterize these cases.  If $\alpha >0.14$,
 the spectral tilt is red.  For example, $n_s = 0.97$ for
$\alpha \approx 2$.   These examples
represent  the range that can be
achieved for the entropically-induced curvature perturbations in
the simplest models, roughly
$0.97 < n_s < 1.02$.

For comparison, if we use the same estimating procedure for the
Newtonian potential fluctuations in the cyclic model (assuming
they converted to curvature fluctuations before the bounce through
5d effects), we obtain $0.95 < n_s < 0.97$. This range agrees with
the estimate obtained  by an independent analysis based on
studying  inflaton potentials directly \cite{Boyle:2005ug}.
Furthermore, as shown in Ref.~\cite{Khoury:2003vb}, the same range
is obtained for time-delay (Newtonian potential) perturbations in
the cyclic model,
 due to a ``duality'' in the perturbations equations~\cite{Boyle:2004gv}.
Hence, all estimates are consistent with one another, and
we can conclude that the range of spectral tilt obtained from
entropically-induced curvature perturbations is typically bluer by
a few percent.

We note that, in the cyclic model, say, it is possible that
curvature perturbations are created both by the entropic mechanism
and by converting Newtonian potential perturbations into curvature
perturbations through 5d effects.  In this case, the
cosmologically relevant contribution is the one with the bigger
amplitude.  In particular, the conversion of Newtonian potential
perturbations is sensitive to the brane collision velocity $V$,
\cite{Tolley:2003nx,McFadden}, whereas the entropic mechanism is
not. So, conceivably, either contribution could dominate.

A second way of analyzing the spectral tilt is to assume a form
for the scalar field potential.
Consider the case
where the two fields have
 steep potentials that can be modelled as
 $V(\phi_1) = -V_0\, e^{-\int c \,\d\phi}$
and $\dot{\phi_2} = \gamma \dot{\phi_1}$. Then
Eq.~(\ref{tilt1}) becomes
\be n_s -1 =
{4(1+\gamma^2) \over c^2 M_{Pl}^2} - { 4 c_{,\phi} \over c^2},
\label{f5} \ee
where we have used the fact that $c(\phi)$ has the dimensions of inverse
mass and restored the factors of Planck mass.  The presence of $M_{Pl}$
 clearly indicates that the first term on
the right is a gravitational term.  It is also the piece that
makes a blue contribution
to the spectral tilt.  The second term is the non-gravitational
term and agrees precisely with the flat space-time
result (\ref{dev}), although the agreement is not at all obvious
at intermediate steps of the calculation.

For a pure exponential potential, which has $c_{,\phi}=0$, the
non-gravitational contribution is zero, and the spectrum is slightly
blue, as our model-independent analysis suggested.  For plausible
values of $c = 20$ and $\gamma = 1/2$, say, the gravitational piece is
about one percent and the spectral tilt is $n_s \approx 1.01$,
also consistent with our earlier estimate.
However, this case with $c_{,\phi}$ precisely equal to
zero is unrealistic.  In the cyclic
model, for example, the steepness of the potential
must decrease as the field rolls downhill in order that the ekpyrotic
phase comes to an end,  which corresponds to $c_{,\phi} >0$.
If $c(\phi)$ changes from some initial value $\bar{c} \gg 1$
to
some value of order unity at the end of the ekpyrotic phase after $\phi$
changes by an amount $\Delta \phi $, then
$c_{,\phi} \sim \bar{c}/\Delta{\phi}$.  When $c$ is large,
the non-gravitational term in
Eq.~(\ref{f5}) typically
dominates and the spectral tilt is a few per cent towards
the red.

For example, suppose $c \propto
\phi^\beta$ and $\int c(\phi)\, \d \phi \approx 125$; then, the
spectral tilt is \be
 n_s-1 = -0.03 {\beta \over 1+\beta}, \label{spectralindex}
\ee which corresponds to $.97 < n_s < 1$ for positive $0<\beta
<\infty$, in agreement with our earlier estimate. We note that
negative potentials of this type with very large values of $c$
have been argued to arise naturally in string theory. For example,
in the work of Conlon and Quevedo \cite{Conlon:2005jm}, the
potential is of the form we require with $c(\phi) \sim C
\phi^{1/3},$ where the constant $C$ is very large. For this
potential, one finds $n_s \approx 0.99$.

Our expression for the spectral tilt of the entropically induced
curvature spectrum can also be expressed in terms of the customary
``fast-roll'' parameters
\cite{Gratton:2003pe} \be \bar{\epsilon}
\equiv \left(\frac{V}{V_{,\phi}}\right)^2 =
\frac{1}{c^2} \qquad \bar{\eta} \equiv
 \Big(\frac{V}{V_{,\phi}}\Big)_{\hspace{-0.5mm},\phi}\,. \ee
Note that $\bar{\epsilon} = 1 /( 2 (1 + \gamma^2) \epsilon)$.
Then, the spectral tilt is
\be
n_{s} -1=
\frac{4(1+\gamma^2)}{M_{Pl}^2}\,\bar{\epsilon} - 4 \bar{\eta}. \ee

This result can be compared with the spectral index of the
time-delay (Newtonian potential) perturbation obtained in earlier
work \cite{Gratton:2003pe}, where the corresponding formula is \be
n_{s}-1 = -\frac{4}{M_{Pl}^2}\, \bar{\epsilon} - 4 \bar{\eta}. \ee
Here, the first term is again gravitational, but it has the
opposite sign of the gravitational contribution to the
entropically induced fluctuation spectrum. So, the tilt is
typically a few per cent redder.

Finally, for inflation, the spectral tilt is \be n_s-1 = -6
\epsilon +2 \eta \ee where the result is expressed in terms of the
slow-roll parameters $\epsilon \equiv (1/2) (M_{Pl}
V_{,\phi}/V)^2$ and $\eta \equiv M_{Pl}^2 V_{,\phi \phi}/V$.  Here
we have revealed the factors of $M_{Pl}$ to illustrate that both
inflationary contributions are gravitational in origin.   This
gives the same range for $n_s$ as the Newtonian potential
perturbations in the cyclic model.

\section{Conclusions}

The entropic mechanism for generating approximately
scale-invariant curvature perturbations in a contracting universe
has two appealing features. First, it can be analyzed entirely
within the context of 4d effective theory. For those who were
skeptical about the ekpyrotic and cyclic models because of their
apparent reliance on 5d effects to create curvature perturbations,
this work shows that there is another, more prosaic mechanism that
can be totally understood in
 familiar terms. This should terminate the debate
on whether it is possible, in principle, to generate curvature
perturbations in a pre-big bang phase.

The second attractive feature is that the essential elements
occur quite naturally in
extra-dimensional theories like string and M-theory. There is
no shortage of scalar field moduli, and, quite generically, these
fields can possess  negative and steeply
decreasing potentials of the ekpyrotic form. In this situation,
approximate scaling solutions exist in which several fields
undergo ekpyrosis simultaneously so that nearly scale-invariant
entropy perturbations are naturally generated. Furthermore, if the
relevant scalar field trajectory encounters a boundary in moduli
space (like that described in Ref. \cite{Lehners:2006ir}), then as
the trajectory reflects off the boundary, entropy perturbations
are naturally  converted into curvature perturbations with
the identical large-scale power spectrum.

We hasten to add that, although we
have only presented here the concrete
example of heterotic M-theory, it is clear that
the present formalism is generic and  can be applied to
other types of  pre-big bang models,
including those that do not rely on there being extra dimensions.

We have also seen that the entropic mechanism has an interesting
signature.  Because of the gravitational contribution to the
spectral tilt of the entropically-induced perturbations, the
spectrum is typically a few per cent bluer than the time-delay
(Newtonian potential) perturbations or the density perturbation in
inflation. To push the inflationary perturbations into this bluer
range requires adding extra degrees of otherwise unnecessary
fine-tuning, as delineated in Ref.~\cite{Boyle:2005ug}.  In
particular, Ref.~\cite{Boyle:2005ug} shows that the natural range
for inflationary models is $0.93< n_s < 0.97$, whereas
entropically-induced spectra tend to lie in a range that is a few
per cent bluer, roughly $0.97<n_s < 1.02$ by our estimates. Hence,
a highly precise measure of the spectral tilt at the one per cent
level or better could serve as an indicator of which mechanism is
responsible. For example, a value of $n_s =0.99$ is awkward to
obtain with inflation but right in the middle of the predicted
range for pre-big bang entropically-induced perturbations.

\begin{center}
***
\end{center}

{\bf Acknowledgements:} We thank Stephen Hawking for discussions.
 JLL and NT are supported by PPARC (UK) and the Center for
Theoretical Cosmology in Cambridge. PLM is supported through the
Dutch Science Organisation (NWO). PJS is supported by
US Department of Energy grant DE-FG02-91ER40671.

\bibliographystyle{apsrev}
\bibliography{ThirdPaper}

\end{document}